\documentclass[12pt]{iopart}
\usepackage{latexsym,amssymb,amsbsy,graphicx}
\usepackage[space]{cite} 
\usepackage{xcolor}
\begin{document}

\title[Time-dependent fluorescence by incoherently pumped polar quantum dot ...
]{Time-dependent fluorescence by incoherently pumped polar quantum
dot driven by
 a low-frequency monochromatic field}


\author{A V Soldatov}

\address{ Department of Mechanics, V.A. Steklov Mathematical
Institute of the Russian Academy of Sciences, 8, Gubkina str.,
Moscow, 119991, Russia} \ead{soldatov@mi-ras.ru}

\begin{abstract}

 We studied time-dependent features of high-frequency fluorescent
radiation from a two-level quantum system with broken inversion
spatial symmetry. The system in question was modelled after a
one-electron two-level asymmetric polar semiconductor quantum dot
whose electric dipole moment operator has permanent unequal
diagonal matrix elements. The dot was permanently excited by
incoherent pumping of some sort. Our attention was focused on the
evolution of the fluorescence spectrum  following abrupt switching
on of an additional  driving monochromatic field, which frequency
is much lower than the optical transition frequency of the quantum
dot.  An analytical expression for the fluorescence spectrum as a
function of the amplitude, initial phase and frequency of the
monochromatic driving field, as well as of the pumping intensity
and the elapsed time, was derived.

\end{abstract}




\markboth{Time-dependent fluorescence by incoherently pumped polar
quantum dot ...}{A.V. Soldatov}

  \maketitle


\newcommand{\noi}{\noindent}
\newcommand{\dint}{\displaystyle\int}
\newcommand{\lam}{\lambda}
\newcommand{\fr}{\frac}
\newcommand{\hb}{\hbar}
\newcommand{\vb}{{\bf b}}
\newcommand{\vba}{{\bf b^+}}
\newcommand{\vp}{{\bf p}}
\newcommand{\vk}{{\bf k}}
\newcommand{\vr}{{\bf r}}
\newcommand{\bk}{{b_\vk}}
\newcommand{\bka}{b^+_\vk}
\newcommand{\akl}{a_{\vk,\lam}}
\newcommand{\akla}{a^+_{\vk,\lam}}
\newcommand{\takl}{\tilde a_{\vk,\lam}}
\newcommand{\takla}{\tilde a^+_{\vk,\lam}}
\newcommand{\ak}{{a_\vk}}
\newcommand{\aka}{{a^+_\vk}}
\newcommand{\om}{\omega}
\newcommand{\Om}{\Omega}
\newcommand{\del}{\delta}
\newcommand{\bg}{\begin{equation}}
\newcommand{\en}{\end{equation}}
\newcommand{\dsum}{\displaystyle\sum}
\newcommand{\nn}{\nonumber}
\newcommand{\lb}{\label}
\newcommand{\lan}{\langle}
\newcommand{\ran}{\rangle}
\newcommand{\Gam}{\Gamma}
\newcommand{\ba}{\begin{eqnarray}}
\newcommand{\ea}{\end{eqnarray}}
\newcommand{\vekl}{\hat\varepsilon_{\vk}^{(\lam)}}
\newcommand{\veps}{\varepsilon}
\newcommand{\sig}{\sigma}
\newcommand{\ah}{{\hat a}}
\newcommand{\ahp}{{\hat a^+}}
\newcommand{\bh}{{\hat b}}
\newcommand{\bhp}{{\hat b^+}}
\newcommand{\hc}{{\hat b}}
\newcommand{\hcp}{{\hat b^+}}
\newcommand{\htap}{\hat{\tilde a}^+}
\newcommand{\hta}{\hat{\tilde a}}
\newcommand{\htsp}{\hat{\tilde S}^+}
\newcommand{\hts}{\hat{\tilde S}^-}

\newcommand{\Del}{\Delta}

\section{Introduction}
\mbox{}\vspace{-\baselineskip}

The study of the spectral properties of electromagnetic radiation
emitted by excited atoms, molecules, and other quantum systems has
always been of paramount interest in various domains of
fundamental and applied quantum optics. As a rule, in most
theoretical and experimental studies it is quite justifiably
assumed that the most common quantum system under study, such as
naturally occurring atoms and molecules, possess the inversion
spatial symmetry. At the same time, examples of violation of this
symmetry are encountered nor so rarely in physical systems of
various nature too. Among such systems one can mention polar
molecules, atomic impurities implanted into the crystal lattice
and located in an asymmetric environment, and Rydberg atoms in an
external asymmetric electrostatic field. All these systems has one
important  feature in common:  in addition to induced transient
electric dipole moments they  may also possess, as a consequence
of the spatial inversion symmetry violation, permanent electric
dipole moments.  This feature increases their options  to interact
with electromagnetic fields. Due to the presence of permanent
electric dipole moments, all such systems are often referred to as
polar systems. By the present time, the variety of such quantum
systems suitable and accessible for experiments and practical
applications has been significantly enlarged due to the discovery
of nontrivial quantum mechanical properties of a new broad class
of quantum systems with finitely many discrete energy levels,
collectively known as quantum dots. Like the natural atoms and
molecules, quantum dots can interact with electromagnetic fields
by means of their induced transient or permanent electric dipole
moments in a wide range of frequencies, which opens up great
opportunities for new refined experiments in the field of quantum
optics and for the use of these quantum mechanical objects to
solve a variety of applied scientific and technological problems,
including problems in the field of opto- and nanoelectronics. As
regards fundamental research, quantum dots provide a unique
opportunity to conduct experiments on single quantum nanoobjects
whose position and orientation are  fixed in the laboratory
coordinate system. Due to their properties, quantum dots are often
called artificial atoms with controlled parameters. Unique optical
properties of polar quantum systems, in which broken inversion
symmetry induces new effects, has already stimulated research
efforts in order to employ them as various radiation sources, see,
e.g. \cite{Izadnenas2023} and refs. [1-12] therein. As a rule,
attention is mostly paid to stationary spectral properties of
these systems. In what follows, non-steady-state spectrum  of the
fluorescence by incoherently pumped one-electron two-level
asymmetric polar semiconductor quantum dot being under the action
of driving low-frequency monochromatic field will  be analyzed.
Stationary fluorescence spectrum of this quantum system was
already studied earlier \cite{lplsold2024}. As was stressed in
\cite{EW2}, there are at least two reasons for studying the
time-dependent fluorescence spectra of quantum systems. The first
one is to study the effect of spectral time-dependence in the
course of spectral measurements by means of a physically realistic
experimental set-up. The second reason is to gain a deeper
understanding of the fluorescence phenomenon itself in various
quantum systems. The present contribution falls in line with this
proclaimed research program.

\section{The model under study}

In what follows, we consider a two-level quantum dot S with a
ground state $|g\ran$, an excited state $|\,e\ran$, the transition
frequency $\om_0$, and the electric dipole moment $\hat{\bf d}$,
being under the action of the external monochromatic field ${\bf
E}(t) ={\bf E}_f \cos(\om_f t+\phi)$ with the amplitude ${\bf
E}_f$ and frequency ${\bf \om}_f$. This two-level system
simultaneously interacts with the dissipative environment, which
can be represented by a thermal reservoir made of a multitude of
vacuum modes of a quantized electromagnetic field. For
convenience, we assume that the shift in the interlevel transition
frequency due to the interaction with the environment is already
taken into account, as usual, in the transition frequency $\om_0$.
Then the Hamiltonian for such a system reads as

 \bg \hspace{-1.0cm}\hat H=\hb\om_0
\hat S^z+ \hb\dsum_j\om_j\hcp_j\hc_j+ \hb\dsum_j\gamma_j(\hcp_j
\hat S^- +\hat S^+\hc_j)-{\bf E}_f\cdot \hat{\bf d}\cos(\om_f
t+\phi),\label{H1}\en

\noi where $\hat S^{\pm}$, $\hat S^z$ - operator variables
belonging to the quantum dot and satisfying conventional
commutation relations

\bg\quad [\hat S^+,\hat S^-]=2 \hat S^z = \hat D, \quad [\hat S^z,
\hat S^{\pm}]= \pm \hat S^{\pm}, \en

\noi   the  Bose operators   $\{\hat b^+_j, \hat b_j\}$ stand for
the creation and annihilation operators of the quantized field
mode quanta with the  frequency $\om_j$, and $\gamma_j$ is the
field-dot interaction constant . The electric dipole moment
operator $\hat{\bf d}$ is defined as

\bg \hat{\bf d} =   {\bf d}_{eg}\hat S^+  +{\bf d}_{ge}\hat S^-
+({\bf d}_{ee}-{\bf d}_{gg})\hat S^z+ \fr{1}{2}({\bf d}_{ee}+{\bf
d}_{gg})\hat I, \en

\bg
 {\bf d}_{eg}=q\langle e|\hat {\bf r}|g\rangle,\quad {\bf
d}_{ge}=q\langle g|\hat {\bf r}|e\rangle,\quad {\bf
d}_{ee}=q\langle e|\hat {\bf r}|e\rangle,\quad {\bf
d}_{gg}=q\langle g|\hat {\bf r}|g\rangle,\en

\noi where $q$ is an electron charge and  $\hat {\bf r}$ is the
operator of the electron position inside the quantum dot. It is
usually assumed that the diagonal matrix elements of the electric
dipole moment operator are equal to zero,
 which is justified for naturally occurring
atoms and numerous species of molecules characterized by definite
parity, so that each of the eigenstates, $|g \rangle$ or $|e
\rangle$, is either symmetric or antisymmetric with respect to the
operation of spatial inversion in a three-dimensional space.
Contrariwise,  artificially fabricated quantum systems, like
asymmetric quantum dots, can possess permanent electric dipole
moments in their ground and excited states due to the spatial
inversion symmetry violation. As a consequence, such dots are
often called polar quantum dots. Hereinafter, we assume that the
quantum dot in the Hamiltonian (\ref{H1}) is polar and, moreover,
that the inequality ${\bf d}_{ee}\ne{\bf d}_{gg}$ holds. In this
case, it is convenient to write the dot-field dipole interaction
term in the form

\bg \hspace{-2.3cm}\!\!-{\bf E_f}\cdot \!\!\hat{\bf d}\cos(\om_s
t+\phi)=\!\hb\!\left[\Om_R S^++{\Om_R}^* S^- + \delta_f^{(as)} S^z
\!-\! \fr{\delta_f^{(sim)}}{2} (|e\rangle\langle e|+
|g\rangle\langle g|)\right]\!\cos(\om_f t+\phi),
 \en

\noi where $\Om_R=-{\bf E}_f\cdot {\bf d}_{eg}/{\hb}$ is the Rabi
frequency, which, without loss of generality, can be taken to be
real and positive, and $\delta_f^{(sim)}={\bf E}_f\cdot({\bf
d}_{gg}+{\bf
 d}_{ee})/\hb$,  $\delta_f^{(as)}={\bf E}_f\cdot({\bf d}_{gg}-{\bf
 d}_{ee})/\hb$\, are the parameters of symmetry violation. The term
proportional to  $\delta_f^{(sim)}$ does not affect the dynamics
of the system and can be omitted.

\section{Evolution equations}

Quantum dynamics of the quantum dot can be described within the
frame of the well-known  Heisenberg-Langevin approach to open
quantum systems \cite{Scully,Gard}. Let us introduce slowly
oscillating operator variables  from the start

\bg \htsp(t)=\hat S^+(t) e^{-i\om_0t},\quad \hts(t)=\hat S^-(t)
e^{i\om_0t} \en

\noi and assume that the frequency of the external monochromatic
driving field,  the parameter $\delta_f^{(as)}$ and the Rabi
frequency $\Om_R$ are much smaller than the optical transition
frequency, i.e. $\om_f,|\,\delta_f^{(as)}|, \Om_R << \om_0$. Under
this assumption one can neglect all rapidly oscillating terms in
favor of slowly oscillating terms in the corresponding initial
complete system of the quantum Heisenberg-Langevin equations
(which complete system is rightly omitted here due to the fact
that all the procedures to be employed to the effect of its
derivation can be found in numerous  textbooks on the theory of
open quantum systems, even with a particular regard to the
problems of quantum optics, e.g. \cite{Scully,Gard}) and arrive at
the system of only two effective relevant operator equations for
slowly oscillating operator variables

\bg \fr{d\hts}{dt}= -\left(\Gam_s +i\del^{(as)}_f\cos(\om_f
t+\phi)\right)\hts +
   \hat F_s(t), \label{GL1}\en

\bg \fr{d\hat D}{dt}= -\Gam_d(\hat D+\hat I)+\Gam_p(\hat I-\hat D)
+ \hat F_d(t), \label{GL2}\en

\noi with  quantum sources of noise which's mean values equal
zero:

\bg
 \hat F_s(t)=i\hat D(t)\dsum_j\gamma_j\hat
b_j(0)e^{-i(\om_j-\om_0)t}, \quad  \lan\hat F_s(t)\ran= 0,\en

\bg \hspace{-1.5cm}\hat F_d(t)=2i\dsum_j\gamma_j(\hat
b^+_j(0)e^{i(\om_j-\om_0)t}\hts(t) - \htsp(t)\hat
b_j(0)e^{-i(\om_j-\om_0)t}),\,\, \lan\hat F_d(t)\ran=0.\en

\noi Here $ \Gam_s = \pi\rho(\om_0)\gamma^2(\om_0),\quad
\tau_s=1/\Gam_s, \quad \Gam_d=1/\tau_d=2\Gam_s, $ where
$\rho(\om_0)$ is the density of the bosonic modes of the
environment in the continuum limit and $\tau_s$, $\tau_d$ are the
characteristic  times of the quantum dot relaxation due to the
interaction with the environment. The incoherent pumping
 of the quantum dot can be accounted for purely phenomehologically
 by  introducing  the term $\Gam_p(\hat I-\hat D)$ into Eq.(\ref{GL2}),
 where the parameter $\Gam_p=1/\tau_p$ stands for the rate of the
 pumping. Consequently,  $ -\Gam_d(\hat D+\hat
I)+\Gam_p(\hat I-\hat D)= -\Gam_D(\hat D -D_0)$, where $
D_0=(\tau_d-\tau_p)/(\tau_d+\tau_p)$ and
$\Gam_D=1/\tau_D=2/\tau_s+1/\tau_p\,. $ Therefore, two independent
equations for the averages $\lan\hts\ran$ and $\lan \hat D\ran$
follow from Eqs.(\ref{GL1}, \ref{GL2}) immediately

\bg \fr{d\lan\hts\ran}{dt}=
-\left(\Gam_s+i\del^{(as)}_f\cos(\om_ft+\phi)\right)\lan\hts\ran,
   \label{av1}  \en

\bg \fr{d\lan \hat D\ran}{dt}= -\Gam_D(\lan\hat D\ran-D_0)
\label{av2}, \en

\noi and the solution to the second of them is

 \bg \lan \hat
D(t)\ran =(\lan \hat D(0)\ran-D_0)e^{-\Gam_D t}
+D_0.\label{sol}\en

\section{Time-dependent high-frequency fluorescence spectrum}

As was proposed in \cite{EW1} and applied in\cite{EW2} for the
study of non-stationary resonance fluorescence, the suitably
normalized counting rate of a photodetector can be used to define
a time-dependent fluorescence spectrum, dubbed the "physical
spectrum" by the authors. This spectrum has the explicit form for
$t\ge 0$

\bg \hspace{-1cm}
S(t,\om,\Gamma)=\fr{\Gamma}{2\pi}\dint_0^tdt_1\dint_0^t dt_2
e^{-(\Gamma/2-i\Delta)(t-t_1)}e^{-(\Gamma/2+i\Delta)(t-t_2)}\lan\hat
{\tilde S}^+(t_1)\hat{\tilde S}^-(t_2)\ran \label{physspect1},\en
 \noi which can be shown to be the same as

\bg\hspace{-1cm}
S(t,\om,\Gamma)=\fr{\Gamma}{\pi}\mbox{Re}\dint_0^tdt_1
e^{-\Gamma(t-t_1)}\dint_0^{t-t_1}d\tau
e^{(\Gamma/2+i\Delta)\tau}\lan\hat {\tilde S}^+(t_1)\hat{\tilde
S}^-(t_1+\tau)\ran \label{physspect2}.\en

\noi Here $t$ is the elapsed time after the already incoherently
pumped for a very long time quantum dot was first subjected to the
low-frequency external field, which means in accordance with
Eq.(\ref{sol}) that $\lan\hat D(t)\ran=D_0$ for all $t\ge 0$, and
$\Gamma$ is the full width of the (effectively Lorentzian)
transmission peak of the Fabry-Perot interferometer while $\Delta$
is the detuning, or frequency offset, of the Fabry-Perot line
centre above the quantum dot transition frequency $\om_0$, i.e.
$\Delta(\om)=\om-\om_0$.

As stated by the quantum regression theorem
\cite{Scully,Gard,Carm}, the correlation function in the integrand
obeys the same temporal evolution equation as  the Eq.(\ref{av1})
for the correspondent averaged operator variable $\lan\hat {\tilde
S}^-(t)\ran$

\bg \hspace{-0.5cm}\fr{d \lan\htsp(t)\hts(t+\tau) \ran}{d\tau}=
-\left(\Gam_s +i\del^{(as)}_f\cos(\om_f(t+\tau)+\phi)\right)
\lan\htsp(t)\hts(t+\tau) \ran \label{corrfeq},
     \en

\noi with  initial condition
 \bg \lan \htsp(t)\hts(t)\ran=  (\lan
\hat D(t)\ran+1)/2\to( D_0+1)/2\en

\noi at $\tau=0$,  which is provided by the asymptote of the
solution (\ref{sol}) to Eq.(\ref{av2}). Herewith it is assumed
that the quantum dot had already been incoherently pumped for a
very long time $t' >> 1/\Gam_D$ before it was first subjected to
the low-frequency external field. Therefore, the solution to
Eq.(\ref{corrfeq}) is given by

\bg \lan \htsp(t)\hts(t+\tau)\ran = \fr{ D_0+1}{2} \times \nn\en
\bg\times
\exp\left(-\Gam_s\tau-i\fr{\delta_f^{(as)}}{\om_f}\sin(\om_f(t+\tau)+\phi)\right)
\exp\left(i\fr{\delta_f^{(as)}}{\om_f}\sin(\om_f t+\phi)\right).
\label{corrf}\en

\noi Now, taking recourse to the Jacobi-Anger identity

\bg \exp\left(i\fr{\delta_f^{(as)}}{\om_f}\sin(\om_f
t+\phi)\right) =
 \dsum_{k=-\infty}^\infty
J_k(\delta_f^{(as)}/\om_f)e^{ik(\om_ft+\phi)}, \en

\noi one can carry out repeated integration in
Eq.(\ref{physspect2}) analytically for arbitrary value of the full
width of the filter transmission peak $\Gamma$:

In what follows, we will consider the case when the quantum dot
had already been incoherently pumped for a long time before been
subjected to low-frequency field at the moment $t=0$. In this case
the stationary  population inversion $\lan\hat D(t\ran)= D_0$ for
all $t\ge 0$, and the spectrum takes the form

\ba S(t,\om,\Gamma)=\fr{\Gamma}{2\pi}(D_0+1)
\mbox{Re}\dsum_{k,p\,=-\infty}^\infty
e^{i(p-k)\phi}J_k(\delta_f^{(as)}/\om_f)J_p(\delta_f^{(as)}/\om_f)e^{-\Gamma
t}\times\nn \ea

\ba\times\left[
\fr{e^{B_{kp}t}-e^{A_{k}t}}{A_kC_p}-\fr{e^{B_{kp}t}-1}{A_kB_{kp}}\right]\label{spectfull},
\ea

\noi where

\ba \hspace{-2cm} A_k=\Gamma/2-\Gamma_s+i(\Del(\om)-k\om_f),\,
B_{kp}=\Gamma+i(p-k)\om_f,\, C_p=
\Gamma/2+\Gamma_s+i(p\om_f-\Del(\om)). \nn
 \ea

\noi It is convenient to rewrite $S(t,\om,\Gamma)$  as

\ba
S(t,\om,\Gamma)=S_{diag}(t,\om,\Gamma)+S_{nondiag}(t,\om,\Gamma),\quad
\ea

\noi where

\ba\hspace{-2.5cm} S_{diag}(t,\om,\Gamma)=\fr{\Gamma}{2\pi}(D_0+1)
\mbox{Re}\dsum_{k=-\infty}^\infty
J^2_k(\delta_f^{(as)}/\om_f)e^{-\Gamma t}\left[
\fr{e^{B_{kk}t}-e^{A_{k}t}}{A_kC_k}-\fr{e^{B_{kk}t}-1}{A_kB_{kk}}\right]=
\ea

\ba=\fr{1}{2\pi}(D_0+1) \dsum_{k=-\infty}^\infty
J^2_k(\delta_f^{(as)}/\om_f)\left[\fr{\Gam_s+\Gam/2}{(\Gam_s+\Gam/2)^2+(\Del(\om)-k\om_f)^2}+\right.\nn
 \ea

 \ba
\hspace{-2.5cm}\left.+ \mbox{Re}\fr{ e^{-\Gamma
t}(\Gam/2+\Gam_s-i(\Del(\om)-k\om_f))-\Gamma
e^{-(\Gam/2+\Gam_s-i(\Del(\om)-k\om_f) )t}
}{(\Gam/2-\Gam_s+i(\Del(\om)-k\om_f))(\Gam/2+\Gam_s-i(\Del(\om)-k\om_f))}\right]
, \ea

\noi and

\ba
\hspace{-2.0cm}S_{nondiag}(t,\om,\Gamma)=\fr{\Gamma}{2\pi}(D_0+1)
\mbox{Re}\dsum_{k,p\, =-\infty\atop{k\ne p}}^\infty
e^{i(p-k)\phi}J_k(\delta_f^{(as)}/\om_f)J_p(\delta_f^{(as)}/\om_f)e^{-\Gamma
t}\times\nn \ea

\ba\times\left[
\fr{e^{B_{kp}t}-e^{A_{k}t}}{A_kC_p}-\fr{e^{B_{kp}t}-1}{A_kB_{kp}}\right]=\nn
\ea

\ba =\fr{\Gamma}{2\pi}(D_0+1) \mbox{Re}\dsum_{k,p\,
=-\infty\atop{k\ne p}}^\infty
e^{i(p-k)\phi}J_k(\delta_f^{(as)}/\om_f)J_p(\delta_f^{(as)}/\om_f)\times\nn
\ea

\ba \hspace{-2.0cm}\times\left[
\fr{e^{i(p-k)\om_ft}}{(\Gam/2+\Gam_s-i(\Del(\om)-p\,\om_f))(\Gam+i(p-k)\om_f)}+\fr{e^{-\Gam
t}}{\Gam/2-\Gam_s+i(\Del(\om)-k\om_f)}\times\right.\nn \ea

\ba
\times\left.\left\{\fr{1}{\Gam+i(p-k)\om_f}-\fr{e^{(\Gam/2-\Gam_s+i(\Del(\om)-k\om_f))t}}{\Gam/2+\Gam_s-i(\Del(\om)-p\,\om_f)}
\right\}\right].\ea

\noi It is worth noticing that the spectrum (\ref{spectfull})
derived above is incoherent due to the fact that the right hand
side of Eq.(\ref{av1}) does not contain any constant term.

\section{Discussion}

It is only natural that the spectrum (\ref{spectfull}) vanishes in
the absence of incoherent pumping, i.e. for $D_0=-1$.  Further on,
the pumping parameter $D_0$, which assumes values between -1 (no
pumping at all) and + 1 (maximum pumping rate resulting in the
complete permanent population inversion of the two-level quantum
dot), was set at the neutral point $D_0=0$, thus providing for the
equally populated ground and excited energy levels of the dot. For
the matter of exemplary numerical calculations, the value of the
damping parameter $\Gam_s=1$, was chosen, and, as follows from the
definitions introduced above, $\tau_D =\tau_s(1-D_0)/4$. In the
limit $\Gam\to 0$, which corresponds to infinitely narrow
transmission peak of the filter,

\bg \lim_{\Gam\to 0}S(t,\om,\Gamma)=0 \en

\noi for any finite elapsed time $t$. Contrariwise, in the
stationary limit of  large time $t$

\ba\hspace{-2cm} \lim_{t\to\infty}S(t,\om,\Gamma)=\fr{D_0+1}{2\pi}
\dsum_{k=-\infty}^\infty
J^2_k(\delta_f^{(as)}/\om_f)\fr{\Gam_s+\Gam/2}{(\Gam_s+\Gam/2)^2+(\Del(\om)-k\om_f)^2}+\nn
 \ea

 \ba+\fr{\Gamma}{2\pi}(D_0+1) \mbox{Re}\dsum_{k,p\,
=-\infty\atop{k\ne p}}^\infty
e^{i(p-k)\phi}J_k(\delta_f^{(as)}/\om_f)J_p(\delta_f^{(as)}/\om_f)\times\nn
\ea

\ba \hspace{-2.0cm}\times\left[
\fr{e^{i(p-k)\om_ft}}{(\Gam/2+\Gam_s-i(\Del(\om)-p\,\om_f))(\Gam+i(p-k)\om_f)}\right].
 \label{fspect3}\ea

\noi In the subsequent limit

\bg \hspace{-2.0cm}S(\om,\Gamma_s)=\lim_{\Gamma\to
0}\lim_{t\to\infty}S(t,\om,\Gamma)= \fr{D_0+1}{2\pi}
\dsum_{k=-\infty}^{\infty}J^2_k(\delta_f^{(as)}/\om_f)\fr{\Gam_s}{\Gam_s^2+(\om-\om_0
-k\om_f)^2}\label{fspect4}\en

\noi the spectrum coincides with the stationary spectrum derived
in \cite{lplsold2024} in the limit of large integration time
$T\to\infty$, where the following theoretical definition for the
spectrum was employed (see, e.g. \cite{Scully}):

\ba S(\om, T)= \fr{1}{2\pi T} \dint_0^T dt'\dint_0^T dt''\lan\hat
{S}^+(t'')\hat { S}^-(t')\ran e^{i\om(t'-t'')}=\nn\ea

\bg= \fr{1}{\pi T}\mbox{Re} \dint_0^T dt'\dint_0^{T-t'}
d\tau\lan\hat {\tilde S}^+(t')\hat {\tilde S}^-(t'+\tau)\ran
e^{i(\om-\om_0)\tau}.\label{spectrum}\en

 \noi An additional broadening of
all the Lorentzian peaks in the first term of the spectrum
(\ref{fspect3}) in comparison with the spectrum (\ref{fspect4}) is
incurred by the filter parameter $\Gamma/2$. So, technically, the
limit $T\to\infty$ in the right hand side of (\ref{spectrum})
 corresponds to the limit $\lim_{\Gam\to
 0}\lim_{t\to\infty}S(t,\om,\Gamma)$. This coincidence shows that
 the perfect theoretically predicted spectrum could  be obtained
 by means of a realistic experimental set-up if only its measurement
 is carried out in the course of infinite elapsed time
  with infinitely narrow filter bandwidth.

   The limiting
stationary spectrum with perfect frequency resolution
(\ref{fspect4}) is composed of a sequence of equidistant
Lorentzian peaks with the half-width $2\Gam_s$ each. All the peaks
are separated by the frequency intervals $\om_f$. The spectrum is
symmetric and centered at the frequency $\om_0$, and all the peaks
are well resolved if $\om_f>>\Gam_s$ (see Figure \ref{figure1}).

 In reality,
the whole picture is a little bit more complicated, simply because
the amplitudes of the peaks are determined by the magnitudes of
$J^2_k(\delta_f^{(as)}/\om_f)$. Therefore,  due to the oscillatory
behavior of the Bessel functions $J_k(x)$, some peaks in the
equidistant sequence can disappear completely for certain values
of $\delta_f^{(as)}/\om_f$, at which points their correspondent
Bessel functions equal zero. Out of the same reason, some peaks in
the spectrum can be much more pronounced than the others. Due to
the limiting behavior  $J^2_k(\delta_f^{(as)}/\om_f)\to0$ of the
Bessel functions for $|\,k|\to\infty$, the discernible spectrum is
bound to concentrate effectively within some finite frequency
interval around the central frequency $\om_0$. As follows from the
numerical analysis,  this interval can be approximately estimated
as $[\om_0-\delta_f^{(as)},\om_0+\delta_f^{(as)}]$ for all
practical purposes.

 In the limit
$\delta_f^{(as)}/\om_f\to 0$ only one single central peak with the
amplitude $(D_0+1)/(2\pi\Gam_s)$ survives at the frequency
$\om_0$. The opposite case $\delta_f^{(as)}/\om_f\to \infty$
implies, actually, two opportunities: either $\delta_f^{(as)}$ is
fixed and $\om_f\to 0$, or, alternatively, $\om_f$ is fixed and
$|\delta_f^{(as)}|\to\infty$. The second opportunity means that,
in reality,  $\om_f<< |\,\delta_f^{(as)}|<< \om_0$, in order to
ensure consistency with the assumption $\om_f,
|\,\delta_f^{(as)}|<< \om_0$ already made above. In the case
$\om_f\to 0$ two tendencies are at play against each other
simultaneously: all the resonance frequencies $\om_0+k\om_f$
gravitate to $\om_0$ while the magnitudes of all the amplitude
coefficients $J^2_k(\delta_f^{(as)}/\om_f)$ gravitate to zero. As
a result of this interplay, the spectrum (\ref{fspect4})  is
taking
 the shape of two symmetric peaks centered at $\om_0\pm
\delta_f^{(as)}$ (compare Figure \ref{figure2} to Figure
\ref{figure3}).

 It is
instructive to pass to the limit $\om_f\to 0$ directly either in
the Hamiltonian (\ref{H1}) or in Eq.(\ref{corrf}). Then, the
correspondent spectrum takes the form of a single peak spectrum
centered at $\om_0+ \delta_f^{(as)}$,

\bg S(\om)\sim \fr{D_0+1}{2\pi} \fr{\Gam_s}{\Gam_s^2+(\om-\om_0
-\delta^{(as)}_f)^2},\label{fspect2}\en

\noi which is obviously different from the double peak limiting
spectrum obtained by direct numerical evaluation of the expression
(\ref{fspect4}) earlier. There is no contradiction, though, simply
because the numerical limiting procedure assumes that the external
monochromatic field keeps oscillating, albeit slower and slower,
all the way down to the final point $\om_f=0$, while in the
opposite case of the spectrum (\ref{fspect2}) this field was
assumed to be constant from the beginning of the calculations.

Under the second opportunity, when $\om_f$ is fixed and
$|\delta_f^{(as)}|\to\infty$, the value of
$J^2_k(\delta_f^{(as)}/\om_f)$ goes to zero for any $k$, the width
of the spectrum increases,  spanning the interval of frequencies
$[\om_0-\delta_f^{(as)},\om_0+\delta_f^{(as)}]$, and eventually
the spectrum becomes, albeit very slowly, less and less
discernible from the base spectral line of frequency (compare
Figure \ref{figure2} to Figure \ref{figure4}).

The major qualitative difference between the two stationary
spectra (\ref{fspect3}) and (\ref{fspect4}) stems from the second
term in Eq.(\ref{fspect3}). This term is proportional to the
filter bandwidth $\Gamma$ and oscillates with the period
$2\pi/\om_f$. These oscillations are always present in the
spectrum but especially pronounced for $\om_f/\delta_f^{(as)}<<1$
(compare Figure (\ref{figure6}) to Figure (\ref{figure7})).
Therefore, this oscillatory behavior is an artifact of the
proposed experimental set-up leading to the definition
(\ref{physspect1}) of the so-called "physical spectrum". At the
same time this phenomenon can be considered as a peculiar optical
effect arising in the compound system made of the physical system
in question and some particular spectrum analyzer.

\noi It is understood, that in a real experimental situation the
phase $\phi$ is most likely cannot be controlled and, therefore,
is random. At the same time, as follows from Eq.(\ref{fspect3}),
the phase may play any role only for some initial time interval
$t\approx 1/\Gamma$ just after the switching on of the
low-frequency field (compare Figure (\ref{figure8}) to Figure
(\ref{figure9}), and also Figure (\ref{figure10})  to  Figure
(\ref{figure11})).   For large enough elapsed time $t>>\Gamma$ the
influence of the phase on the spectrum only results in its
temporal phase shift .

 The relations $\Gamma_d=2 \Gamma_s$ and
$\Gam_D=2\Gamma_s+1/\tau_p$ between the model parameters  may
create an impression that the validity  of this study is limited
to low temperatures exclusively, because usually the polarization
dephasing rate ($\Gamma_s \sim$ THz) in quantum dots is much
larger than the population relaxation rate ($\Gamma_D \sim$ GHz)
at normal conditions. In reality, this impression is rather
misleading. This spurious dependence between $\Gamma_s$ and
$\Gamma_d$ is an artifact stipulated solely by the specific choice
of that term in the Hamiltonian (\ref{H1}) responsible for the
interaction with the environment. This term does not account for
any dephasing mechanisms. This unnecessary drawback can be readily
ameliorated by incorporation of  additional terms of the proper
kind into the model Hamiltonian. Eventually, one will arrive at
the same set of equations (\ref{av1}) and (\ref{av2}), but with
totally independent parameters $\Gamma_s$ and $\Gamma_d$.
Therefore, it is still justified to apply the results of this
study to the case when $\Gamma_s>>\Gamma_D$ and see exactly the
same structure of spectra caused by the low-frequency field
modulation.

\section{Summary}

To summarize, we used the counting-rate spectrum definition
proposed in \cite{EW1} to study the transient and asymptotic
features of the fluorescence radiation spectrum emitted by
incoherently pumped one-electron two-level asymmetric polar
semiconductor quantum dot being under the action of driving
low-frequency monochromatic field. An analytical expression for
this spectrum as a function of the amplitude, initial phase and
frequency of the monochromatic driving field,  bandwidth of the
Fabry-Perot interferometer filter transmission peak, as well as of
the incoherent pumping intensity and the time elapsed after the
switching on of the driving field, was derived. It was shown that
the smaller the filter bandwidth $\Gamma$ the longer time it takes
for the stationary fluorescence spectrum to develop. As to the
influence of the initial phase $\phi$ of the monochromatic driving
field on the spectral characteristics, it may alter the spectrum
for some time after the moment of the monochromatic field
introduction but, finally, its influence only results in the
temporal phase shift of the spectrum in the long run. The most
interesting, albeit counterintuitive, result is the steady
temporal dependence of the so-called "physical spectrum"
(\ref{fspect3}) on the elapsed time $t$ in the stationary limit
$t\to\infty$. In this limit the spectrum shows undamped periodical
oscillations with the period of the driving monochromatic field
$2\pi/\om_f$. A degree to which this effect might be noticeable
and observable  depends on the values of the driving field
parameters $\om_f$, $\delta^{(as)_f}$ and the bandwidth of the
filter $\Gamma$. To some extent, this effect, be it happen to be
non-existent at all in reality, may pose a question about the
validity of the whole "physical spectrum" concept.


\begin{figure}[th]
\vspace{0.0cm}\hspace{7pc}
\begin{minipage}{22.5pc}
\includegraphics[width=22.5pc]{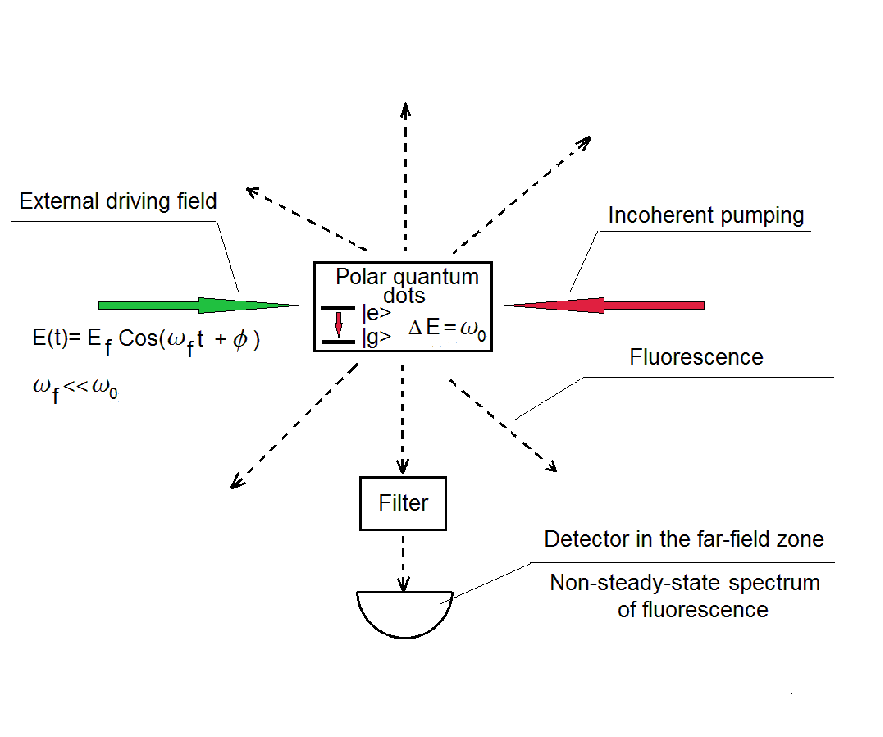}\vspace{-0.5pc}
\caption{Schematic set-up of non-steady fluorescence
experiment}\label{f2tor}
\end{minipage}
\end{figure}

\begin{figure}[th]
\begin{minipage}{18.4pc}
\includegraphics[width=18.4pc]{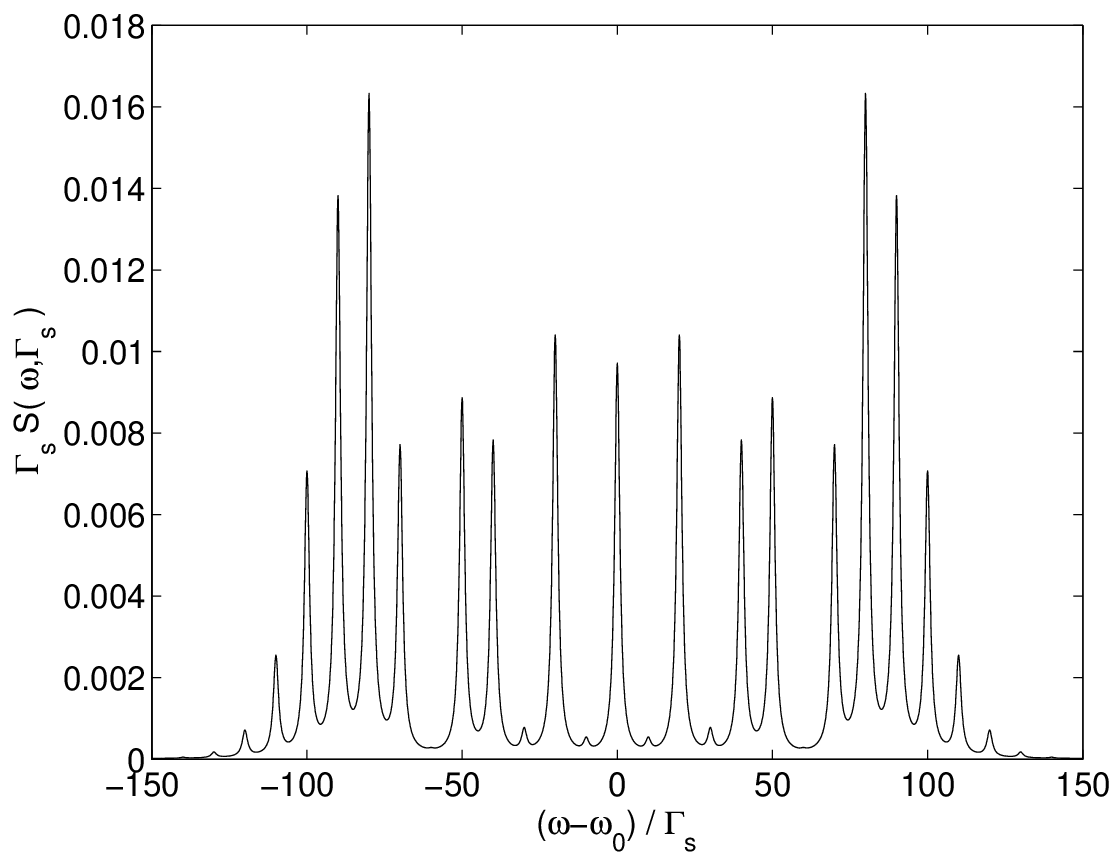}
\caption{\label{figure1} HF spectrum \newline $\Gamma_s=1$,
$\delta^{(as)}_f=100\Gam_s$, $\omega_f=10\Gam_s$ }
\end{minipage}\hspace{1.0pc}%
\begin{minipage}{18.4pc}\vspace{-0.3pc}
\includegraphics[width=18.4pc]{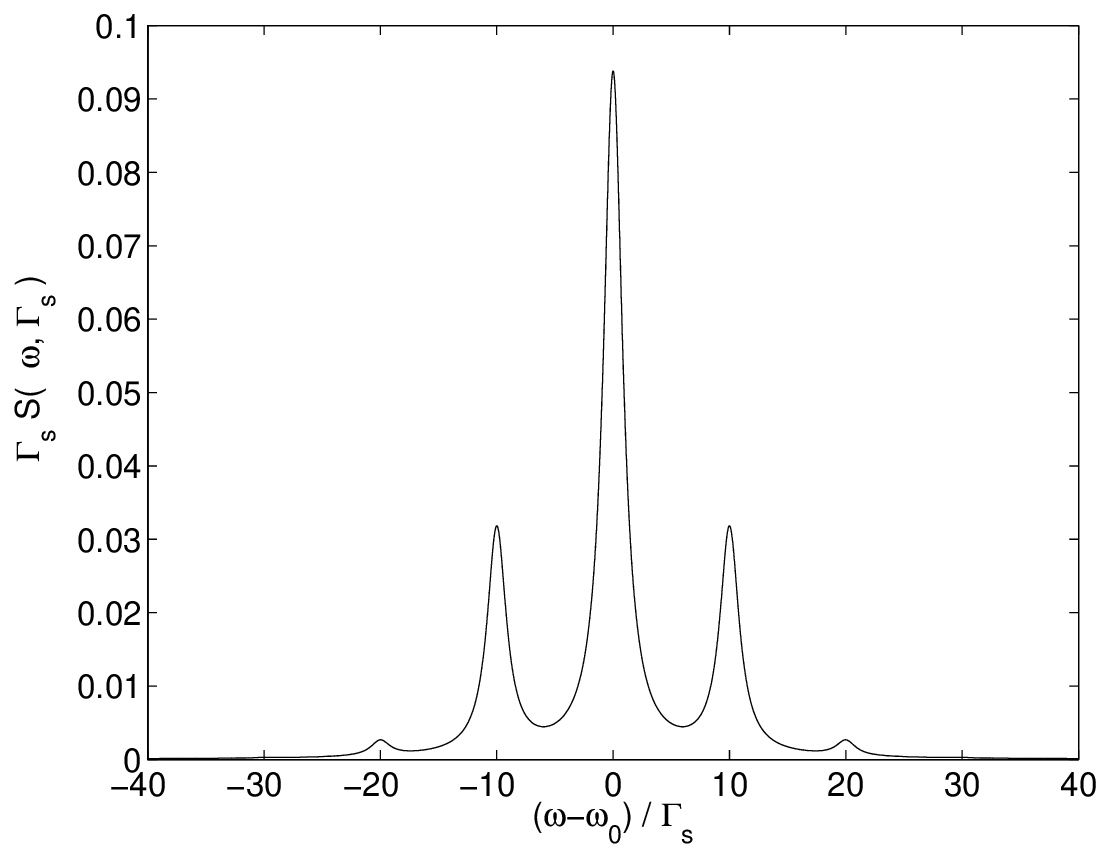}\vspace{-0.5pc}
\caption{\label{figure2}HF spectrum \newline $\Gamma_s=1$,
$\delta^{(as)}_f=10\Gam_s$, $\omega_f=10\Gam_s$ }
\end{minipage}
\end{figure}

\begin{figure}[th]
\begin{minipage}{18.4pc}
\includegraphics[width=18.4pc]{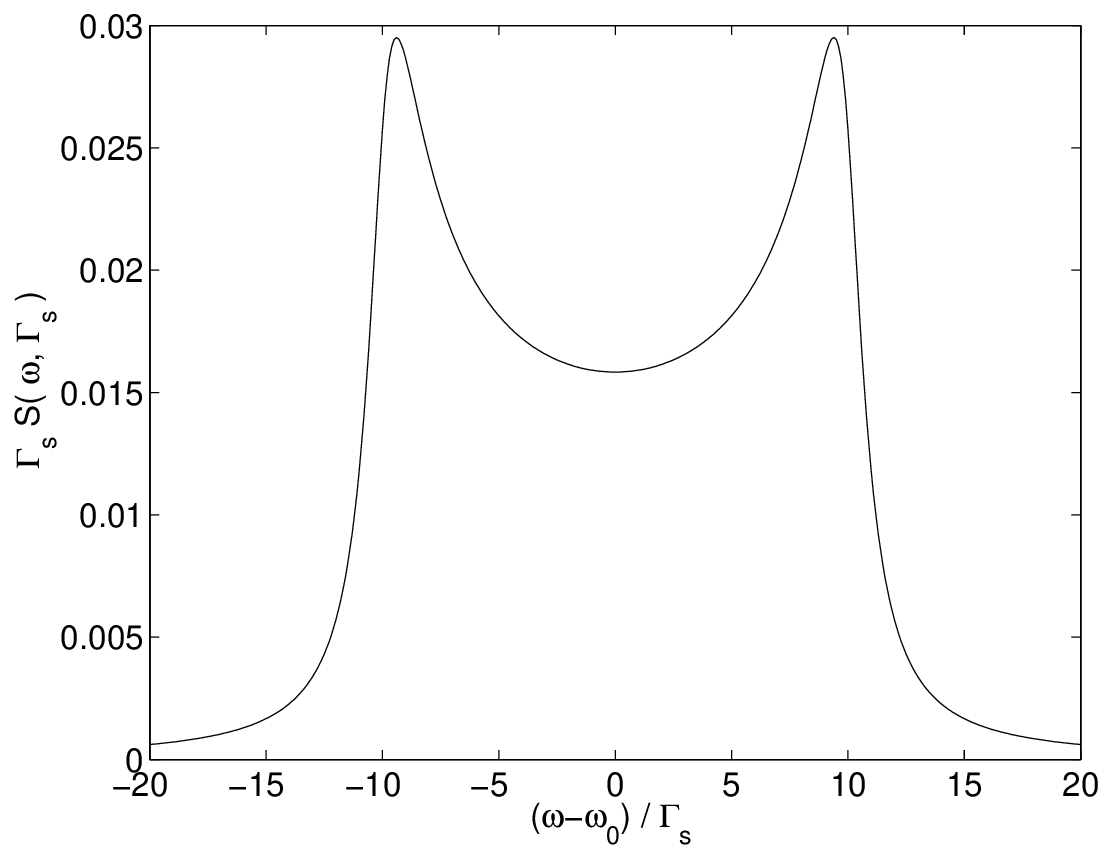}
\caption{\label{figure3} HF spectrum \newline $\Gamma_s=1$,
$\delta^{(as)}_f=10\Gam_s$, $\omega_f=0.1\Gam_s$ }
\end{minipage}\hspace{1.0pc}%
\begin{minipage}{18.4pc}\vspace{-0.3pc}
\includegraphics[width=18.4pc]{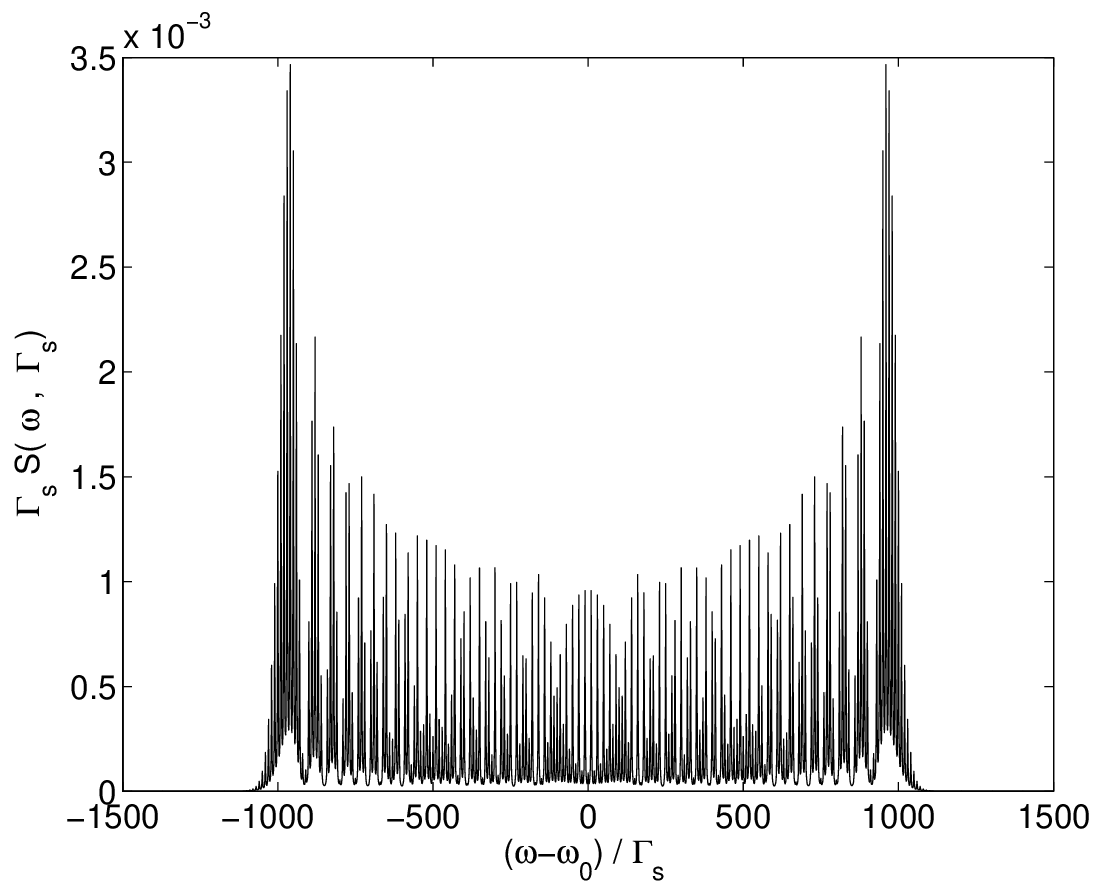}\vspace{-0.5pc}
\caption{\label{figure4}HF spectrum \newline $\Gamma_s=1$,
$\delta^{(as)}_f=1000\Gam_s$, $\omega_f=10\Gam_s$ }
\end{minipage}
\end{figure}

\begin{figure}[th]
\begin{minipage}{18.4pc}
\includegraphics[width=18.4pc]{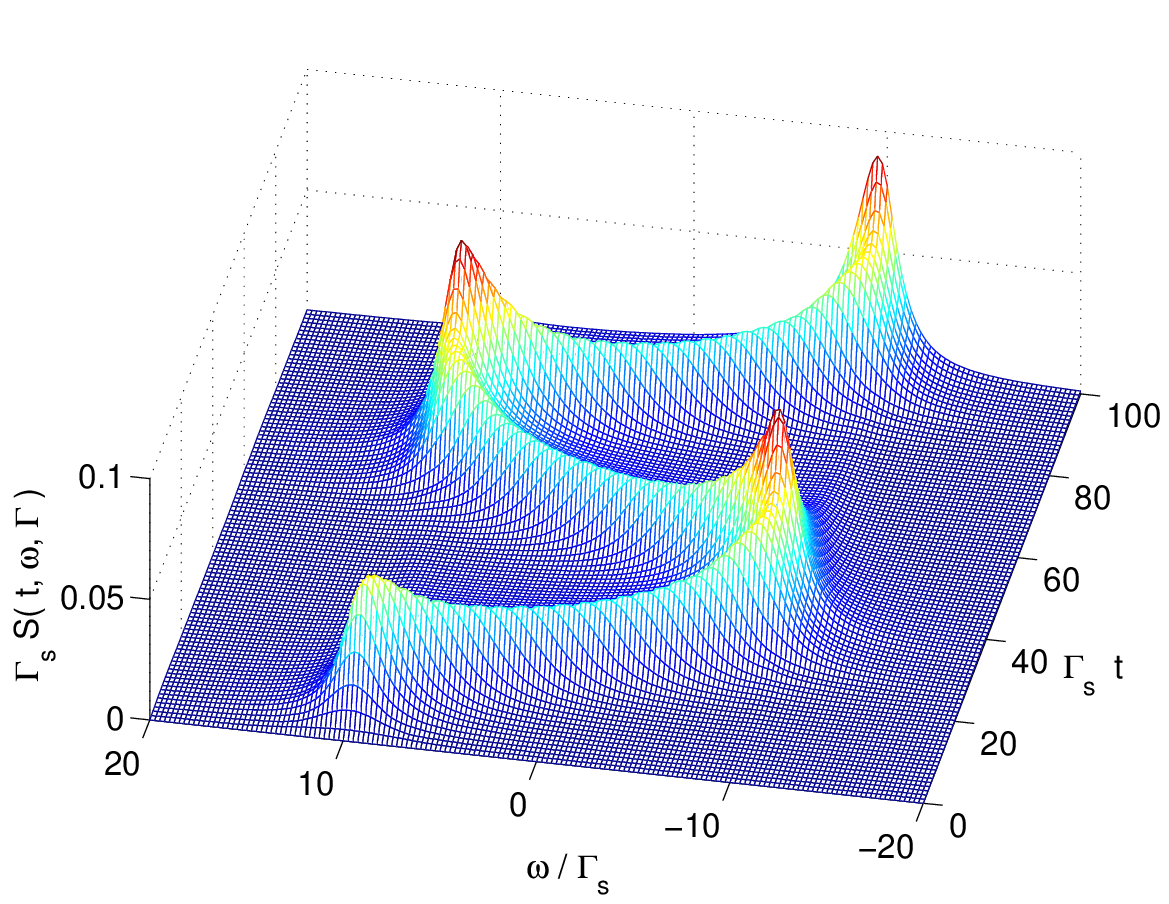}
\caption{\label{figure6} HF spectrum \newline $\Gamma_s=1$,
$\Gamma=0.1 \Gamma_s$, \, $\delta^{(as)}_f=10\Gam_s$,
$\omega_f=0.1\Gam_s$, $\phi=0$, $\Gamma_s t=0:1:100$ }
\end{minipage}\hspace{1.0pc}%
\begin{minipage}{18.4pc}\vspace{-0.3pc}
\includegraphics[width=18.4pc]{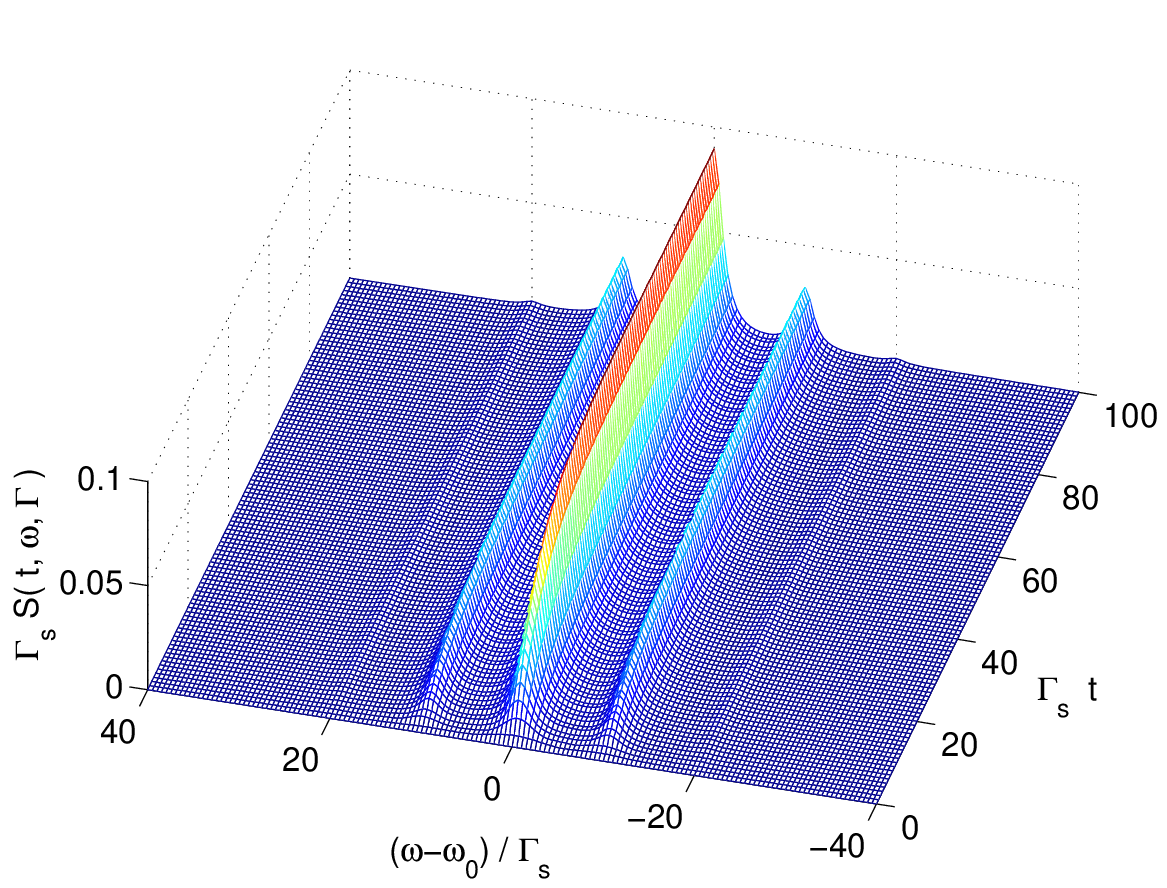}\vspace{-0.5pc}
\caption{\label{figure7}HF spectrum \newline $\Gamma_s=1$,
$\Gamma=0.1 \Gamma_s$, \,$\delta^{(as)}_f=10\Gam_s$,
$\omega_f=10\Gam_s$, $\phi=0$, $\Gamma_s t=0:1:100$ }
\end{minipage}
\end{figure}

\begin{figure}[th]
\begin{minipage}{18.4pc}
\includegraphics[width=18.4pc]{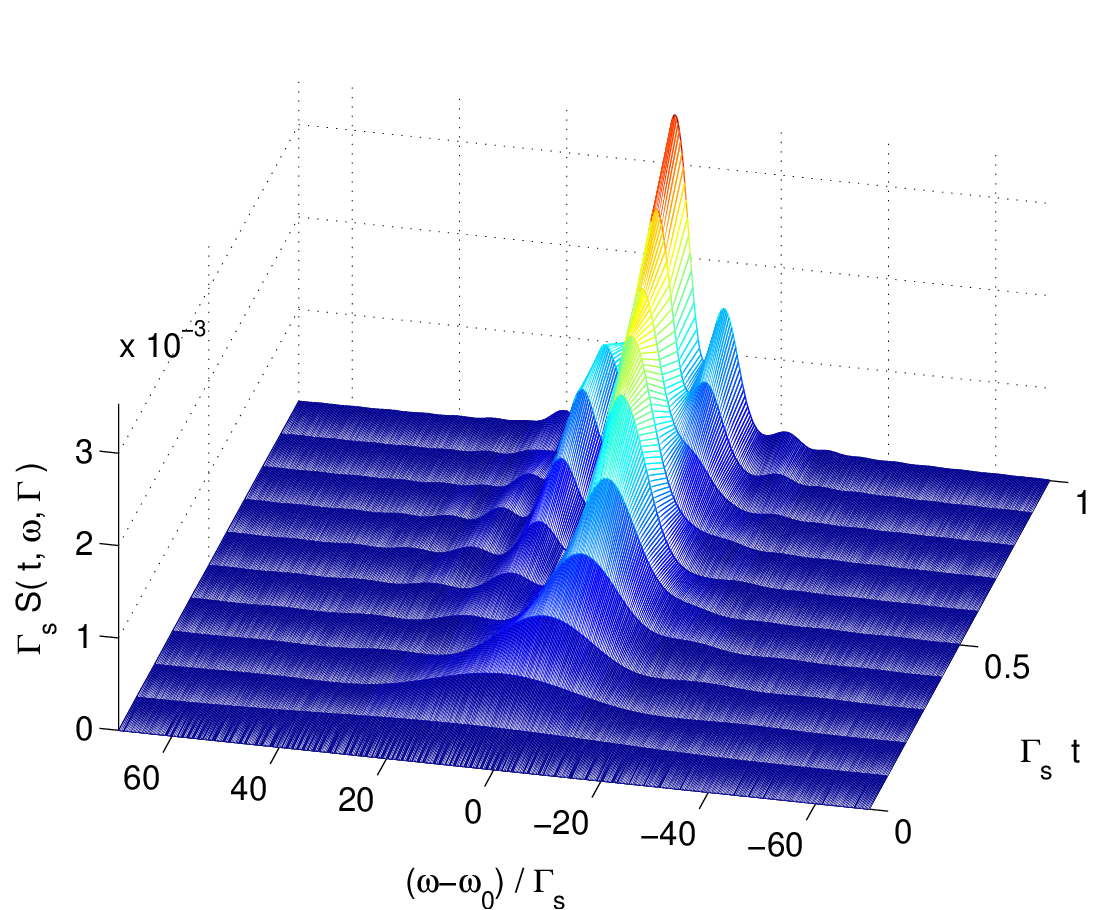}
\caption{\label{figure8} HF spectrum \newline $\Gamma_s=1$,
$\Gamma=0.1 \Gamma_s$, \,$\delta^{(as)}_f=10\Gam_s$,
$\omega_f=10\Gam_s$, $\phi=0$, $\Gamma_s t=0:0.1:1$ }
\end{minipage}\hspace{1.0pc}%
\begin{minipage}{18.4pc}\vspace{-0.3pc}
\includegraphics[width=18.4pc]{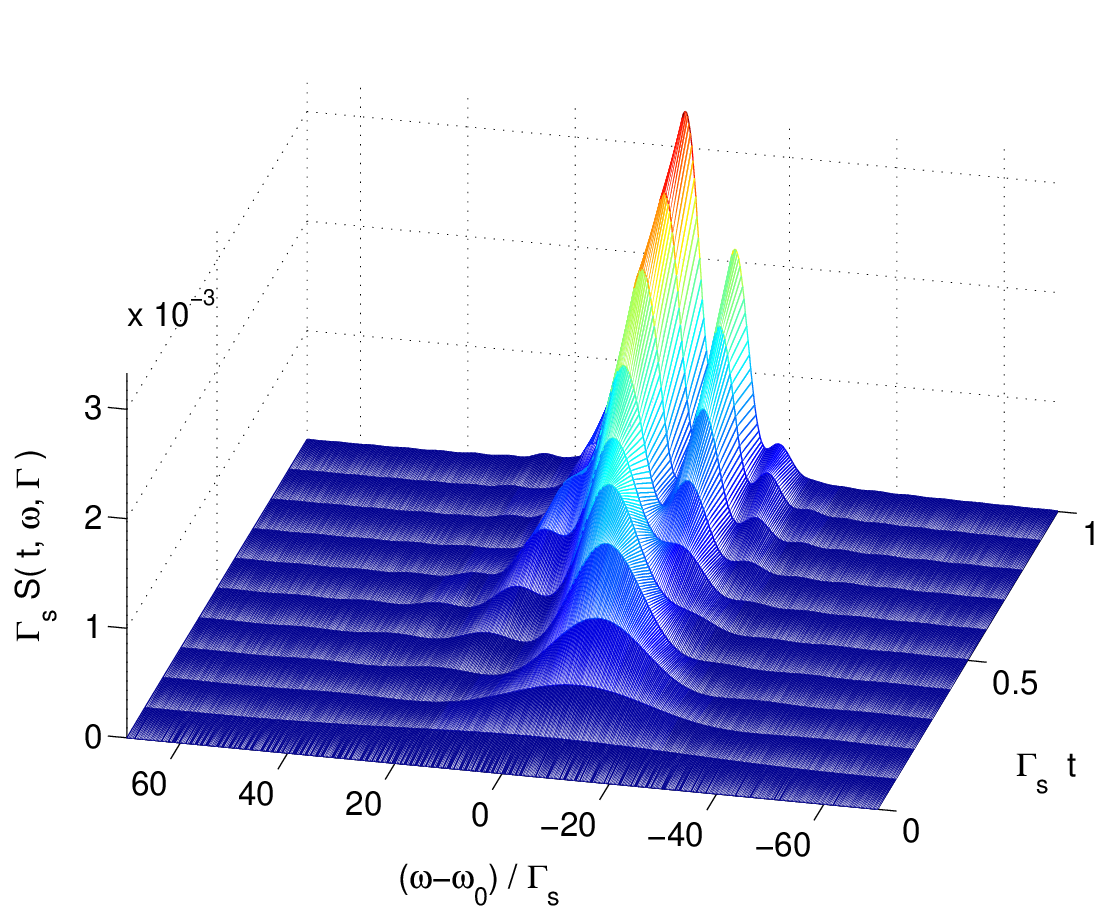}\vspace{-0.5pc}
\caption{\label{figure9}HF spectrum \newline $\Gamma_s=1$,
$\Gamma=0.1 \Gamma_s$, \,$\delta^{(as)}_f=10\Gam_s$,
$\omega_f=10\Gam_s$, $\phi=\pi/2$, $\Gamma_s t=0:0.1:1$ }
\end{minipage}
\end{figure}

\begin{figure}[th]
\begin{minipage}{18.4pc}
\includegraphics[width=18.4pc]{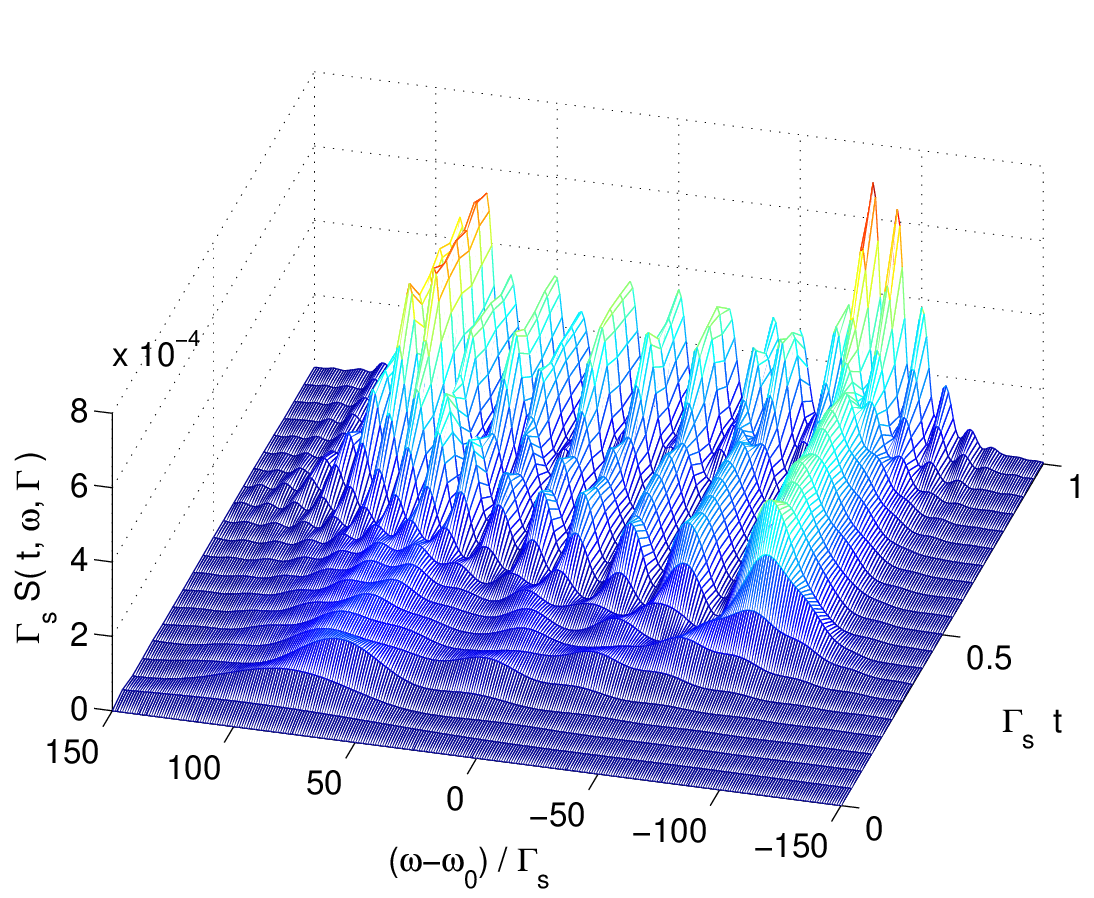}
\caption{\label{figure10} HF spectrum \newline $\Gamma_s=1$,
$\Gamma=0.1 \Gamma_s$, \,$\delta^{(as)}_f=100\Gam_s$,
$\omega_f=10\Gam_s$, $\phi=0$, $\Gamma_s t=0:0.05:1$ }
\end{minipage}\hspace{1.0pc}%
\begin{minipage}{18.4pc}\vspace{-0.3pc}
\includegraphics[width=18.4pc]{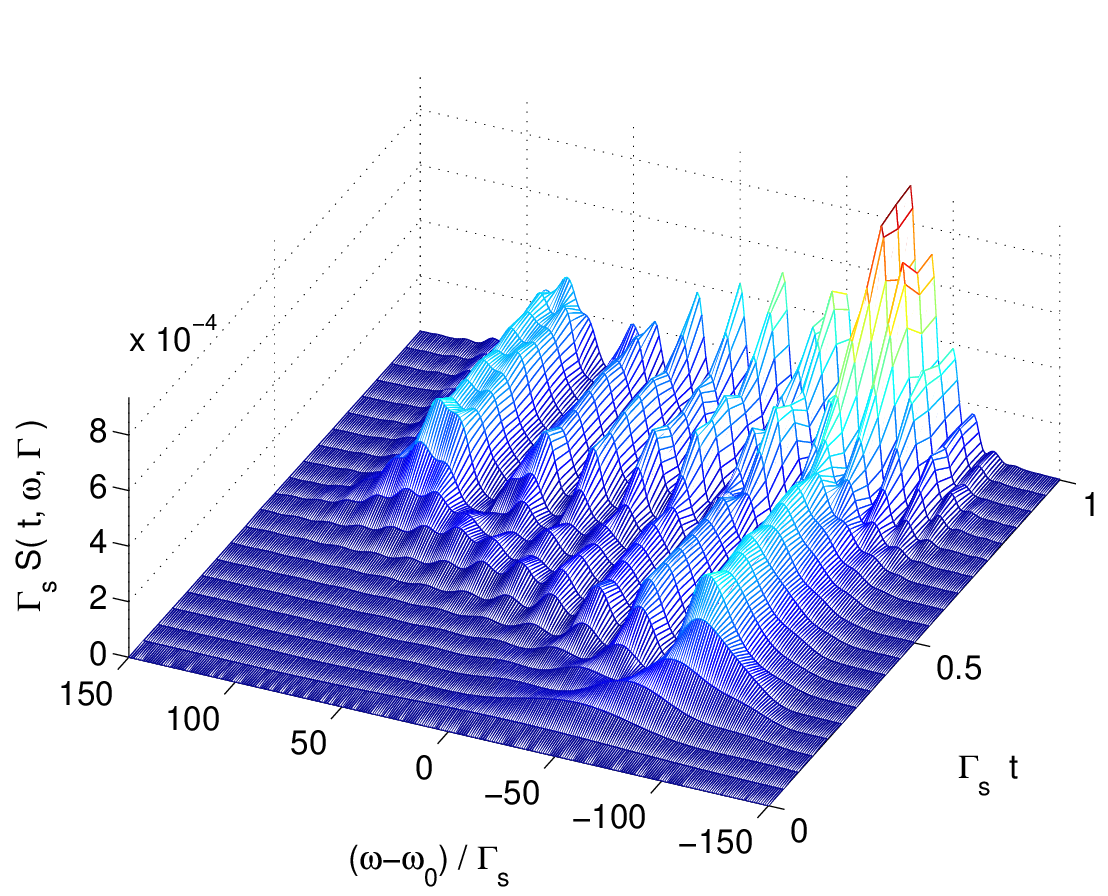}\vspace{-0.5pc}
\caption{\label{figure11}HF spectrum \newline $\Gamma_s=1$,
$\Gamma=0.1 \Gamma_s$, \,$\delta^{(as)}_f=100\Gam_s$,
$\omega_f=10\Gam_s$, $\phi=\pi/2$, $\Gamma_s t=0:0.05:1$ }
\end{minipage}
\end{figure}

\end {document}